\documentclass [6pt,a4paper]{article}
\usepackage{bbm}
\usepackage{amsfonts}
\usepackage{mathrsfs}
\usepackage {amssymb}
\usepackage {amsmath}
\usepackage{amsthm}
\usepackage{latexsym}
\def\dse#1{\vskip 0.6cm\noindent
        {\large\bf #1}
        \vskip 0.4cm}
\newcommand{\pf}{{\bf Proof. \ }}
\def\dse#1{\vskip 0.6cm\noindent
        {\large\bf #1}
        \vskip 0.4cm}
\usepackage{lastpage}
\usepackage{epsfig}

\begin{document}
\begin{center}
\textbf{\large{On the Construction of Optimal Asymmetric Quantum
Codes}}\footnote { E-mail addresses: liqiwangg@163.com(L.Wang),
zhushixin@hfut.edu.cn(S.Zhu).}
\end{center}

\begin{center}
{Liqi Wang, Shixin Zhu}
\end{center}

\begin{center}
\textit{School of Mathematics, Hefei University of
Technology Hefei 230009, Anhui, P.R.China}
\end{center}

\noindent\textbf{}Constacyclic codes are important classes of linear codes that have been applied to the construction of quantum codes. Six new families of asymmetric quantum codes derived from constacyclic codes are constructed in this paper. Moreover, the constructed asymmetric quantum codes are optimal and different from the codes available in the literature. \\

\noindent\emph{Keywords}: Quantum codes; Asymmetric quantum codes; constacyclic codes

\dse{1~~Introduction} Quantum error-correcting codes have gained
prominence since the initial discovery of Shor$^1$ and Steane$^2$.
In 1998, Calderbank et al.$^3$ presented systematic methods to construct binary quantum codes, called stabilizer codes or additive codes, from classical error-correcting codes. Since then the field has made rapid progress, many good
binary quantum codes were constructed by using classical
error-correcting codes, such as BCH codes, Reed-Solomon codes,
Reed-Muller codes, and algebraic geometric codes (see Refs. 4-8).
The theory was later extended to the nonbinary case, since the
realization that nonbinary quantum codes can use fault-tolerant
quantum computation (see Refs. 9-13). Recently, a number of new types of quantum codes, such as convolutional quantum codes, subsystem quantum codes have been studied and the stabilizer method has been extended to these variations of quantum codes (see Refs. 14-15).

Asymmetric quantum error-correcting codes(AQECC) are quantum codes defined over quantum channels where qudit-flip errors and phase-shift errors may have different probabilities. AQECC was first studied by Steane in [16]. Since then, the construction of quantum codes have extended to asymmetric quantum channels. Loffe et al.$^{17}$ utilize BCH codes to correct qubit-flip errors and LDPC codes to correct more frequently phase-shift errors. AQECC derived from LDPC codes and BCH codes were also constructed in [18-21]. Stephens et al.$^{22}$ consider the investigation of AQECC via code conversion. Wang et al.$^{23}$ presented the construction of nonadditive AQECC as well as constructions of asymptotically good AQECC derived from algebraic-geometry codes. Ezerman et al.$^{24}$ presented the construction of AQECC under the trace Hermitian inner product. Ezerman and Ling$^{25}$ studied two systematic construction of AQECC. Chee et al.$^{26}$ constructed pure $q$-ary AQECC and some of these codes attain the quantum Singleton bound. Recently, Ezerman et al.$^{27}$ also studied the pure AQECC and some optimal codes are obtained. A variety of the constructions of new AQECC were presented in [28-31].

AQECC attain the quantum Singleton bound are called optimal. Until now, just several families of optimal AQECC have been constructed. Chee et al.$^{26}$ constructed optimal AQECC with parameters $[[2^m+2, 2^m-4, 4/4]]_{2^m}$ using generalized RS codes, where $m$ is a positive integer. Guardia $^{28}$ constructed optimal AQECC with parameters $[[p-1, p-2d+2, d/(d-1)]]_p$, where $p$ is a prime number. Qian$^{32}$ constructed optimal AQECC with parameters $[[q^2+1, q^2+1-2(k+i+2), (2k+3)/(2i+3)]]_{q^2}$, where $0\leq k\leq i\leq q/2-1$. Recently, Chen et al.$^{33}$ constructed two families of optimal AQECC derived from negacyclic codes. In this paper, we constructed six new families of optimal AQECC derived from constacyclic codes. They are given by

\begin{description}
\item (1) $[[n,n-s-t,(s+1)/(t+1)]]_{q^2},$ where $n=(q^2-1)/2$ and $1\leq t\leq s\leq q-1$.
\item (2) $[[n,n-s-t,(s+1)/(t+1)]]_{q^2},$ where $n=\lambda(q-1)$, $\lambda=(q+1)/r$, $r\neq 2$ is an even divisor of $q+1$ and $1\leq t\leq s\leq (q-1)/2.$
\item (3) $[[n,n-s-t,(s+1)/(t+1)]]_{q^2},$ where $n=\lambda(q+1)$, $\lambda$ is an odd divisor of $q-1$ and $1\leq t\leq s\leq (q-1)/2+\lambda.$
\item (4) $[[n,n-s-t,(s+1)/(t+1)]]_{q^2},$ where $n=2\lambda(q+1)$, $\lambda$ is an odd divisor of $q-1$, $q\ \equiv\ 1\ mod\ 4$, and $1\leq t\leq s\leq (q-1)/2+2\lambda.$
\item (5) $[[n, n-2(s+t+1), (2s+2)/(2t+2)]]_{q^2}$, where $n=(q^2+1)/5$, $q=20m+3$ or $20m+7$ with $m$ a positive integer, and $0\leq t\leq s\leq (q+1)/4.$
\item (6) $[[n, n-2(s+t+1), (2s+2)/(2t+2)]]_{q^2}$, where $n=(q^2+1)/5$, $q=20m-3$ or $20m-7$ with $m$ a positive integer, and $0\leq t\leq s\leq (q-1)/4.$
\end{description}

The paper is organized as follows. In Section 2, some definitions and basic results
of constacyclic codes are reviewed. In Section 3, we recall some basic definitions of asymmetric quantum codes.
In Section 4, six classes of optimal asymmetric quantum codes are constructed. Section 5 concludes the paper.

\dse{2~~Review of Constacyclic Codes}
Let $\mathbb{F}_{q^2}$ be the Galois field with
$q^2$ elements, where $q$ is a power of a prime $p$. A linear $[n, k]$
code $C$ over $\mathbb{F}_{q^2}$ is a $k$-dimensional subspace of
$\mathbb{F}_{q^2}^n$. A linear code $C$ of length $n$ over
$\mathbb{F}_{q^2}$ is called $\eta$-constacyclic if it is invariant under the $\eta$-constacyclic
shift of $\mathbb{F}_{q^2}^n$:
$$(c_0,c_1,\ldots,c_{n-1})\rightarrow(\eta c_{n-1},c_0,\ldots,c_{n-2}),$$
where $\eta$ is a nonzero element of $\mathbb{F}_{q^2}$.
Each codeword $\textbf{c}=(c_0,c_1,\ldots,c_{n-1})$ is customarily identified
with its polynomial representation
$c(x)=c_0+c_1x+\cdots+c_{n-1}x^{n-1}$, and the code $C$ is in turn
identified with the set of all polynomial representations of its
codewords. Then in the ring $\frac{\mathbb{F}_{q^2}[x]}{\langle
x^n-\eta\rangle},$ $xc(x)$ corresponds to a $\eta$-constacyclic shift of $c(x)$. It
is well known that a linear code $C$ of length $n$ over
$\mathbb{F}_{q^2}$ is $\eta$-constacyclic if and only if $C$ is an ideal of the
quotient ring $\frac{\mathbb{F}_{q^2}[x]}{\langle x^n-\eta\rangle}.$
Moreover, $\frac{\mathbb{F}_{q^2}[x]}{\langle x^n-\eta\rangle}$ is a
principal ideal ring, whose ideals are generated by monic factors of
$x^n-\eta$, i.e., $C=\langle f(x)\rangle$ and $f(x)|(x^n-\eta)$.

The Hermitian inner product is defined as $$\langle \textbf{x}, \textbf{y}\rangle=x_0\bar{y}_0+x_1\bar{y}_1+\cdots+x_{n-1}\bar{y}_{n-1}\in\mathbb{F}_{q^2},$$
where $\textbf{x}=(x_0, x_1,\ldots,x_{n-1})\in\mathbb{F}_{q^2}^n$, $\textbf{y}=(y_0, y_1,\ldots,y_{n-1})\in\mathbb{F}_{q^2}^n$, and $\bar{y}_i=y_i^q$.
The vectors $\textbf{x}$ and $\textbf{y}$ are called orthogonal with respect to the Hermitian inner product if $\langle \textbf{x}, \textbf{y}\rangle=0.$ For a $q^2$-ary linear code $C$ of length $n$, the Hermitian dual code of $C$ is defined as $$C^{\bot_H}=\{\textbf{x}\in\mathbb{F}_{q^2}^n|\langle\textbf{x}, \textbf{y}\rangle=0\ for\ all\  \textbf{y}\in C\}.$$
A linear code $C$ of length $n$ over $\mathbb{F}_{q^2}$ is called Hermitian self-orthogonal if $C\subseteq C^{\bot_H}$, and
it is called Hermitian self-dual if $C=C^{\bot_H}.$

We assume $\gcd(q, n)=1$. Let $\delta$ be a primitive $rn$th root of unity in some extension field of $\mathbb{F}_{q^2}$ such that $\delta^n=\eta$. Let $\xi=\delta^r,$ then $\xi$ is a primitive $n$th root of unity. Hence,  $$x^n-\eta=\Pi_{i=0}^{n-1}(x-\delta\xi^i)=\Pi_{i=0}^{n-1}(x-\delta^{1+ir}).$$

Let $\Omega=\{1+ir|0\leq i\leq n-1\}.$ For each $j\in\Omega,$ let $C_j$ be the $q^2$-cyclotomic coset modulo $rn$ containing $j$. Let $C$ be an $\eta$-constacyclic code of length $n$ over $\mathbb{F}_{q^2}$ with generator polynomial $g(x)$. Then the set $Z=\{j\in\Omega|g(\delta^j)=0\}$ is called the defining set of $C$. It is clearly to see the defining set of $C$ is a union of some $q^2$-cyclotomic cosets modulo $rn$ and $dim(C)=n-|Z|$. It is also easily to see $C^{\bot_H}$ has defining set $Z^{\bot_H}=\{z\in\Omega|-qz\ mod\ rn\not\in Z\}$(See Ref. 13).\\

The following results in [34, 35] play an important role in constructing asymmetric quantum codes.\\

\noindent\textbf{Theorem 2.1} \emph{ (The BCH bound for constacyclic codes) Assume that $\gcd(q, n)=1$. Let $C=\langle g(x)\rangle$ be an $\eta$-constacyclic code of length $n$ over $\mathbb{F}_{q^2}$ with the roots $\{\delta^{1+ir}|0\leq i\leq d-2\}$, where $\delta$ is a primitive $rn$th root of unity. Then the minimum distance of $C$ is at least $d$.}\\

\noindent\textbf{Proposition 2.2} \emph{(Singleton bound) Let $C$ be an $[n,k,d]$ linear code over $\mathbb{F}_{q^2}$, then $k\leq n-d+1.$}\\

\noindent\textbf{Lemma 2.3} \emph{ Let $C_i$ be an $\eta$-constacyclic code of length $n$ over $\mathbb{F}_{q^2}$ with defining set $Z_i$ for $i=1,2$. Then $C_1\subseteq C_2$ if and only if $Z_2\subseteq Z_1.$ }

%\noindent\textbf{Lemma 2.4} \emph{ Let $C$ be an $\eta$-constacyclic code of length $n$ over $\mathbb{F}_{q^2}$ with defining set $Z\subseteq\Omega$. Then $C$ contains its Hermitian dual code if and only if $Z\cap Z^{-q}=\emptyset,$ where $Z^{-q}=\{-qz\ mod\ rn|z\in Z\}$.}

\dse{3~~Asymmetric Quantum Codes}

In this section, we recall some basic definitions and results of asymmetric quantum codes. More details we refer to [10, 28-31].

Let $\mathbb{H}$ be the Hilbert space
$\mathbb{H}=\mathbb{C}^{q^n}=\mathbb{C}^q\otimes\cdots\otimes\mathbb{C}^q,$ where $\mathbb{C}^q$ denotes a $q$-dimensional complex vector space representing the states of a quantum mechanical system.
Let $|x\rangle$ be the vectors of an orthonormal basis of
$\mathbb{C}^q$, where the labels $x$ are elements of $\mathbb{F}_q.$
Let $a, b$ be two elements of $\mathbb{F}_q.$ The unitary operators
$X(a)$ and $Z(b)$ on $\mathbb{C}^q$ are defined as
$X(a)|x\rangle=|x+a\rangle$ and
$Z(b)|x\rangle=\omega^{tr(bx)}|x\rangle,$ respectively, where $tr$
is the trace map from $\mathbb{F}_q$ to the prime field $\mathbb{F}_p$ and
$\omega=\textrm{exp}(2\pi i/p)$ is a primitive $p$th root of unity.
Let $\textbf{a}=(a_1, a_2,\ldots,a_n)\in\mathbb{F}_q^n$ and $\textbf{b}=(b_1, b_2,\ldots,b_n)\in\mathbb{F}_q^n.$ Denote
$X(\textbf{a})=X(a_1)\otimes X(a_2)\otimes\cdots\otimes X(a_n)$ and
$Z(\textbf{b})=Z(b_1)\otimes Z(b_2)\otimes\cdots\otimes Z(b_n)$ by
the tensor products of $n$ error operators. The set
$\varepsilon_n=\{X(\textbf{a})Z(\textbf{b})|\textbf{a},\textbf{b}\in\mathbb{F}_q^n\}$
is an error basis on $\mathbb{C}^q$ and the set
$G_n=\{\omega^cX(\textbf{a})Z(\textbf{b})|\textbf{a},\textbf{b}\in\mathbb{F}_q^n,\
c\in\mathbb{F}_p\}$ is the error group associated with
$\varepsilon_n.$

For a quantum error $e=\omega^cX(\textbf{a})Z(\textbf{b})\in G_n,$ the quantum weight $w_Q(e),$ the $X$-weight $w_X(e)$ and the $Z$-weight $w_Z(e)$ of $e$, are defined respectively as
$$w_Q(e)=\sharp\{i: 1\leq i\leq n, (a_i, b_i)\neq(0, 0)\},$$
$$w_X(e)=\sharp\{i: 1\leq i\leq n, a_i\neq 0\},$$
$$w_Z(e)=\sharp\{i: 1\leq i\leq n, b_i\neq 0\}.$$

\noindent\textbf{Definition 3.1}[10,15,16,20] \emph{ A $q$-ary asymmetric quantum code $C$, denoted by $[[n, k, d_z/d_x]]_q$, is a $q^k$-dimensional subspace of the Hilbert space $\mathbb{H}$ and corrects all qudit-flip errors up to $\lfloor\frac{d_x-1}{2}\rfloor$ and all phase-shift errors up to $\lfloor\frac{d_z-1}{2}\rfloor$.}\\

The following well-known CSS construction shown in [18-20] will be utilized in this paper:\\

\noindent\textbf{Theorem 3.2} \emph{ Let $C_i$ be a classical linear code with parameters $[n, k_i, d_i]_{q^2}$, for $i=1,2$. If $C_1^{\bot_H}\subseteq C_2$, then there exists an asymmetric quantum code with parameters $[[n, k_2+k_1-n, d_z/d_x]]_{q^2}$, where $d_z=wt(C_2\backslash C_1^{\bot_H})$ and $d_x=wt(C_1\backslash C_2^{\bot_H})$.}\\

For a CSS asymmetric quantum codes $[[n, k, d_z/d_x]]_{q^2}$, the relations among $n,k, d_z$ and $d_x$ have the following famous result:\\

\noindent\textbf{Theorem 3.3} \emph{ If a CSS asymmetric quantum code $C$ with parameters $[[n, k, d_z/d_x]]_{q^2}$ exists, then $C$ satisfies the asymmetric quantum Singleton bound $$k\leq n-d_z-d_x+2.$$ Especially, if $k=n-d_z-d_x+2,$ then $C$ is called an optimal code.}

\dse{4~~Code Construction}

In this section, we construct six classes of asymmetric quantum codes  based
on constacyclic codes over
$\mathbb{F}_{q^2}.$

\subsubsection*{4.1~~Construction I}

In this section we construct asymmetric quantum codes from constacyclic codes of length $n=\lambda(q-1)$ with $\lambda$ a divisor of $q+1$ over $\mathbb{F}_{q^2}$, where the classical codes are endowed with the Hermitian inner product. Let $r=(q+1)/\gcd(v,q-1)$ be even, for some $v\in\{1,2,\ldots,q\}$. Let $\eta=\omega^{v(q-1)}$ and $\lambda=(q+1)/r$. For each $0\leq j\leq n-1,$ note that the $q^2$-cyclotomic coset containing $1+jr$ modulo $rn$ has only one element $1+jr$. The following Lemma from [13] plays an important role in the asymmetric quantum codes construction.\\

\noindent\textbf{Lemma 4.1} [13, Lemma 3.1]\emph{ Let $r=(q+1)/\gcd(v,q-1)$ be even, for some $v\in\{1,2,\ldots,q\}$. Let $n=\lambda(q-1)$ with $\lambda=(q+1)/r.$ Suppose that $C$ is an $\eta$-constacyclic code of length $n$ over $\mathbb{F}_{q^2}$ with defining set $Z=\bigcup_{j=1}^{\delta}C_{1+r(j-1)}.$ Then
\begin{description}
\item 1) if $r=2$ and $1\leq \delta\leq q-1$, then $C^{\bot_H}\subseteq C$;
\item 2) if $r\neq2$ and $1\leq \delta\leq (q-1)/2$, then $C^{\bot_H}\subseteq C$.
\end{description}}

If $r=2$, then $n=(q^2-1)/2$ and $C$ is a negacyclic code over $\mathbb{F}_{q^2}$. Now we give the first construction of this paper:\\

\noindent\textbf{Theorem 4.2} \emph{ Let $q$ be an odd prime power, and $n=(q^2-1)/2$. Then there exist asymmetric quantum codes with parameters $[[n, n-s-t, (s+1)/(t+1)]]_{q^2}$, where $s, t$ are positive integers and $1\leq t\leq s\leq q-1.$}\\

\noindent\pf Suppose $C_2$ is a negacyclic code over $\mathbb{F}_{q^2}$ of length $n$ with defining set $Z_2=\bigcup_{i=1}^{t}C_{2i-1}$, where $1\leq t\leq q-1$. Then the dimension of $C_2$ is $n-t$. Observe that $Z_2$ consists of $t$ consecutive odd integers $\{1, 3, \ldots, 2t-1\}$. From the BCH bound for constacyclic codes, the minimum distance of $C_2$ is at least $t+1$. From Proposition 2.2, we can see that the minimum distance of $C_2$ is $t+1$. Hence, $C_2$ is a negacyclic code with parameters $[n, n-t,t+1]_{q^2}$.

Now, suppose $C_1$ is a negacyclic code over $\mathbb{F}_{q^2}$ of length $n$ with defining set $Z_1=\bigcup_{i=1}^{s}C_{2i-1}$, where $1\leq t\leq s\leq q-1$. By Lemma 4.1, $C_1^{\bot_H}\subseteq C_1$ and the dimension of $C_1$ is $n-s$. Observe that $Z_2$ consists of $s$ consecutive odd integers $\{1, 3, \ldots, 2s-1\}$. From the BCH bound for constacyclic codes, the minimum distance of $C_1$ is at least $s+1$. From Proposition 2.2, we get the minimum distance of $C_1$ is $s+1$. Hence, $C_1$ is a negacyclic code with parameters $[n, n-s,s+1]_{q^2}$. It is clearly to see that $C_1^{\bot_H}\subseteq C_2$. Then from Theorem 3.2, there exist asymmetric quantum codes with parameters $[[n, n-s-t, (s+1)/(t+1)]]_{q^2}.$ \qed\\

If $r\neq 2$, we can get the following asymmetric quantum codes:\\

\noindent\textbf{Theorem 4.3} \emph{ Let $r\neq 2$ be an even divisor of $q+1$. Let $n=\lambda(q-1)$ with $\lambda=(q+1)/r$. Then there exist asymmetric quantum codes with parameters $[[n, n-s-t, (s+1)/(t+1)]]_{q^2}$, where $s, t$ are positive integers and $1\leq t\leq s\leq (q-1)/2.$}\\

\noindent\pf Let $\eta=\omega^{\lambda(q-1)}$, where $\omega$ is a primitive element of $\mathbb{F}_{q^2}.$ Let $C_2$ be the $\eta$-constacyclic code over $\mathbb{F}_{q^2}$ of length $n$ with defining set $Z_2=\bigcup_{i=1}^{t}C_{1+r(i-1)}$, where $1\leq t\leq (q-1)/2$. Then the dimension of $C_2$ is $n-t$. Observe that $Z_2$ consists of $t$ odd integers $\{1, 1+r, 1+2r, \ldots, 1+(t-1)r\}$. The minimum distance of $C_2$ is at least $t+1$ from Theorem 2.1. Furthermore, we can see that the minimum distance of $C_2$ is $t+1$ from Proposition 2.2. Hence, $C_2$ is an $\eta$-constacyclic code with parameters $[n, n-t,t+1]_{q^2}$.

Now, suppose $C_1$ is an $\eta$-constacyclic code over $\mathbb{F}_{q^2}$ of length $n$ with defining set $Z_1=\bigcup_{i=1}^{s}C_{1+r(i-1)}$, where $1\leq t\leq s\leq (q-1)/2$. Similar to the discussion of $C_2$, $C_1$ is an $\eta$-constacyclic code with parameters $[n, n-s,s+1]_{q^2}$. It is clearly to see that $C_1^{\bot_H}\subseteq C_2$ by Lemma 2.3 and Lemma 4.1. Then from Theorem 3.2, there exist asymmetric quantum codes with parameters $[[n, n-s-t, (s+1)/(t+1)]]_{q^2}.$ \qed\\

\noindent\textbf{Example 4.4} {Let $q=9$, then $n=(q^2-1)/2=40$. Applying Theorem 4.2 produces asymmetric quantum codes in Table 1.}

\begin{table}[htbp]
{\small
\begin{center}
{\small{\bf TABLE 1}~~Asymmetric quantum codes derived from constacyclic codes of
length 40\\}
\begin{tabular}{c c c c c c c c c c c c c c c c c c c}
\hline
& $[[40,38,2/2]]_{81}$& & &$[[40,35,4/3]]_{81}$ & & &$[[40,31,7/4]]_{81}$& &  &$[[40,29,7/6]]_{81}$&\\
& $[[40,37,3/2]]_{81}$& & &$[[40,34,5/3]]_{81}$ & & &$[[40,30,8/4]]_{81}$ & & &$[[40,28,8/6]]_{81}$&\\
& $[[40,36,4/2]]_{81}$& & &$[[40,33,6/3]]_{81}$ & & &$[[40,29,9/4]]_{81}$& & &$[[40,27,9/6]]_{81}$&\\
& $[[40,35,5/2]]_{81}$& & &$[[40,32,7/3]]_{81}$& & &$[[40,32,5/5]]_{81}$& & &$[[40,28,7/7]]_{81}$&\\
& $[[40,34,6/2]]_{81}$& & &$[[40,31,8/3]]_{81}$& & &$[[40,31,6/5]]_{81}$& & &$[[40,27,8/7]]_{81}$&\\
& $[[40,33,7/2]]_{81}$& & &$[[40,30,9/3]]_{81}$& & &$[[40,30,7/5]]_{81}$& & &$[[40,26,9/7]]_{81}$&\\
& $[[40,32,8/2]]_{81}$& & &$[[40,34,4/4]]_{81}$& & &$[[40,29,8/5]]_{81}$& & &$[[40,26,8/8]]_{81}$&\\
& $[[40,31,9/2]]_{81}$& & &$[[40,33,5/4]]_{81}$& & &$[[40,28,9/5]]_{81}$& & &$[[40,25,9/8]]_{81}$&\\
& $[[40,36,3/3]]_{81}$& & &$[[40,32,6/4]]_{81}$& & &$[[40,30,6/6]]_{81}$& & &$[[40,24,9/9]]_{81}$&\\
\hline
\end{tabular}
\end{center}}
\end{table}

\noindent\textbf{Example 4.5} {Let $q=11$ and $r=4$, then $\lambda=3$ and $n=\lambda(q-1)=30$. Applying Theorem 4.3 produces asymmetric quantum codes in Table 2.}

\begin{table}[htbp]
{\small
\begin{center}
{\small{\bf TABLE 2}~~Asymmetric quantum codes derived from constacyclic codes of
length 30\\}
\begin{tabular}{c c c c c c c c c c c c c c c c c c c}
\hline
& $[[30,28,2/2]]_{121}$& & &$[[30,26,3/3]]_{121}$ & & &$[[30,23,5/4]]_{121}$&\\
& $[[30,27,3/2]]_{121}$& & &$[[30,25,4/3]]_{121}$ & & &$[[30,22,6/4]]_{121}$&\\
& $[[30,26,4/2]]_{121}$& & &$[[30,24,5/3]]_{121}$ & & &$[[30,22,5/5]]_{121}$&\\
& $[[30,25,5/2]]_{121}$& & &$[[30,23,6/3]]_{121}$& & &$[[30,21,6/5]]_{121}$&\\
& $[[30,24,6/2]]_{121}$& & &$[[30,24,4/4]]_{121}$& & &$[[30,20,6/6]]_{121}$&\\
\hline
\end{tabular}
\end{center}}
\end{table}

\subsubsection*{4.2~~Construction II}

In this section we construct asymmetric quantum codes from constacyclic codes of length $n=\lambda(q+1)$ with $\lambda$ an odd divisor of $q-1$ over $\mathbb{F}_{q^2}$, where the classical codes are endowed with the Hermitian inner product. Let $v=(q+1)/2$, then $\eta=-1$. It is easy to see that $2n$ divides $q^2-1$. Hence, for each odd $i$ in the range $1\leq i\leq 2n$, the $q^2$-cyclotomic coset $C_i$ modulo $2n$ is $C_i=\{i\}$.\\

\noindent\textbf{Lemma 4.6} [13, Lemma 3.6]\emph{ Let $n=\lambda(q+1)$, where $\lambda$ an odd divisor of $q-1$. If $C$ is a $q^2$-ary negacyclic code of length $n$ with defining set $Z=\bigcup_{j=1}^{\delta}C_{2j-1},$ where $1\leq\delta\leq(q-1)/2+\lambda$, then $C^{\bot_H}\subseteq C$.}\\

\noindent\textbf{Theorem 4.7} \emph{ Let $n=\lambda(q+1)$ with $\lambda$ an odd divisor of $q-1$. Then there exist asymmetric quantum codes with parameters $[[n, n-s-t, (s+1)/(t+1)]]_{q^2}$, where $s, t$ are positive integers and $1\leq t\leq s\leq (q-1)/2+\lambda.$}\\

\noindent\pf Suppose $C_2$ is a negacyclic code over $\mathbb{F}_{q^2}$ of length $n$ with defining set $Z_2=\bigcup_{i=1}^{t}C_{2i-1}$, where $1\leq t\leq (q-1)/2+\lambda$. Then the dimension of $C_2$ is $n-t$. Observe that $Z_2$ consists of $t$ consecutive odd integers $\{1, 3, \ldots, 2t-1\}$. From Theorem 2.1, the minimum distance of $C_2$ is at least $t+1$. From Proposition 2.2, we can see that the minimum distance of $C_2$ is $t+1$. Hence, $C_2$ is a negacyclic code with parameters $[n, n-t,t+1]_{q^2}$.

Now, suppose $C_1$ is a negacyclic code over $\mathbb{F}_{q^2}$ of length $n$ with defining set $Z_1=\bigcup_{i=1}^{s}C_{2i-1}$, where $1\leq t\leq s\leq (q-1)/2+\lambda$. Similar to the discussion of $C_2$, $C_1$ is a negacyclic code with parameters $[n, n-s,s+1]_{q^2}$. Then from Lemma 2.3 and Theorem 3.2, there exist asymmetric quantum codes with parameters $[[n, n-s-t, (s+1)/(t+1)]]_{q^2}.$ \qed\\

\noindent\textbf{Lemma 4.8} [13, Lemma 3.9]\emph{ Assume that $q\equiv 1\ mod\ 4$. Let $n=2\lambda(q+1)$, where $\lambda$ is an odd divisor of $q-1$. If $C$ is a $q^2$-ary negacyclic code of length $n$ with defining set $Z=\bigcup_{j=1}^\delta C_{2j-1}$, where $1\leq \delta\leq(q-1)/2+2\lambda$, then $C^{\bot_H}\subseteq C$.}\\

\noindent\textbf{Theorem 4.9} \emph{Let $q\equiv 1\ mod\ 4$ and $n=2\lambda(q+1)$ with $\lambda$ an odd divisor of $q-1$. Then there exist asymmetric quantum codes with parameters $[[n, n-s-t, (s+1)/(t+1)]]_{q^2}$, where $s, t$ are positive integers and $1\leq t\leq s\leq (q-1)/2+2\lambda.$}\\

\noindent\pf Suppose $C_2$ is a negacyclic code over $\mathbb{F}_{q^2}$ of length $n$ with defining set $Z_2=\bigcup_{i=1}^{t}C_{2i-1}$, where $1\leq t\leq (q-1)/2+2\lambda$. Then the dimension of $C_2$ is $n-t$. Observe that $Z_2$ consists of $t$ consecutive odd integers $\{1, 3, \ldots, 2t-1\}$. From Theorem 2.1, the minimum distance of $C_2$ is at least $t+1$. From Proposition 2.2, we can see that the minimum distance of $C_2$ is $t+1$. Hence, $C_2$ is a negacyclic code with parameters $[n, n-t,t+1]_{q^2}$.

Now, suppose $C_1$ is a negacyclic code over $\mathbb{F}_{q^2}$ of length $n$ with defining set $Z_1=\bigcup_{i=1}^{s}C_{2i-1}$, where $1\leq t\leq s\leq (q-1)/2+\lambda$. Similar to the discussion of $C_2$, $C_1$ is a negacyclic code with parameters $[n, n-s,s+1]_{q^2}$. Then from Lemma 2.3 and Theorem 3.2, there exist asymmetric quantum codes with parameters $[[n, n-s-t, (s+1)/(t+1)]]_{q^2}.$ \qed\\

\noindent\textbf{Remark} From Theorem 4.2, Theorem 4.3, Theorem 4.7 and Theorem 4.9, $d_z+d_x=s+t+2$. Then from Theorem 3.3, the constructed asymmetric quantum codes with parameters $[[n, n-s-t, (s+1)/(t+1)]]_{q^2}$ attain asymmetric quantum Singleton bound. Hence, these asymmetric quantum codes are optimal.\\

\noindent\textbf{Example 4.10} {Let $q=7$ and $\lambda=3$, then $n=\lambda(q+1)=24$. Applying Theorem 4.7 produces asymmetric quantum codes in Table 3.}

\begin{table}[htbp]
{\small
\begin{center}
{\small{\bf TABLE 3}~~Asymmetric quantum codes derived from constacyclic codes of
length 24\\}
\begin{tabular}{c c c c c c c c c c c c c c c c c c c}
\hline
& $[[24,22,2/2]]_{49}$& & &$[[24,19,4/3]]_{49}$ & & &$[[24,15,7/4]]_{49}$&\\
& $[[24,21,3/2]]_{49}$& & &$[[24,18,5/3]]_{49}$ & & &$[[24,16,5/5]]_{49}$&\\
& $[[24,20,4/2]]_{49}$& & &$[[24,17,6/3]]_{49}$ & & &$[[24,15,6/5]]_{49}$&\\
& $[[24,19,5/2]]_{49}$& & &$[[24,16,7/3]]_{49}$& & &$[[24,14,7/5]]_{49}$&\\
& $[[24,18,6/2]]_{49}$& & &$[[24,18,4/4]]_{49}$& & &$[[24,14,6/6]]_{49}$&\\
& $[[24,17,7/2]]_{49}$& & &$[[24,17,5/4]]_{49}$& & &$[[24,13,7/6]]_{49}$&\\
& $[[24,20,3/3]]_{49}$& & &$[[24,16,6/4]]_{49}$& & &$[[24,12,7/7]]_{49}$&\\
\hline
\end{tabular}
\end{center}}
\end{table}

\noindent\textbf{Example 4.11} {Let $q=9$ and $\lambda=1$, then $n=2\lambda(q+1)=20$. Applying Theorem 4.9 produces asymmetric quantum codes in Table 4.}

\begin{table}[htbp]
{\small
\begin{center}
{\small{\bf TABLE 4}~~Asymmetric quantum codes derived from constacyclic codes of
length 20\\}
\begin{tabular}{c c c c c c c c c c c c c c c c c c c}
\hline
& $[[20,18,2/2]]_{81}$& & &$[[20,15,4/3]]_{81}$ & & &$[[20,11,7/4]]_{81}$&\\
& $[[20,17,3/2]]_{81}$& & &$[[20,14,5/3]]_{81}$ & & &$[[20,12,5/5]]_{81}$&\\
& $[[20,16,4/2]]_{81}$& & &$[[20,13,6/3]]_{81}$ & & &$[[20,11,6/5]]_{81}$&\\
& $[[20,15,5/2]]_{81}$& & &$[[20,12,7/3]]_{81}$& & &$[[20,10,7/5]]_{81}$&\\
& $[[20,14,6/2]]_{81}$& & &$[[20,14,4/4]]_{81}$& & &$[[20,10,6/6]]_{81}$&\\
& $[[20,13,7/2]]_{81}$& & &$[[20,13,5/4]]_{81}$& & &$[[20,9,7/6]]_{81}$&\\
& $[[20,16,3/3]]_{81}$& & &$[[20,12,6/4]]_{81}$& & &$[[20,8,7/7]]_{81}$&\\
\hline
\end{tabular}
\end{center}}
\end{table}

\subsubsection*{4.3~~Construction III}

In this section, we construct asymmetric quantum codes with a special length $n=(q^2+1)/5$ over $\mathbb{F}_{q^2}$. \\

\noindent\textbf{Lemma 4.12} [13, Lemma 3.12]\emph{ Let $n=(q^2+1)/5$ and $k=(q^2+1)/2$. Then, for any integer $i\in\Omega=\{1+(q+1)j|0\leq j\leq n-1\},$ the $q^2$-cyclotomic coset $C_i$ modulo $(q+1)n$ is given by
\begin{description}
\item 1) $C_k=\{k\}$ and $C_{k+n(q+1)/2}=\{k+n(q+1)/2\}.$
\item 2) $C_{k-(q+1)j}=\{k-(q+1)j, k+(q+1)j\}$ for $1\leq j\leq n/2-1.$
\end{description}}

\noindent\textbf{Lemma 4.13} [13, Lemma 3.13]\emph{ Let $q$ be an odd prime power with the form $20m+3$ or $20m+7$, where $m$ is a positive integer. Let $n=(q^2+1)/5$ and $k=(q^2+1)/2$. If $C$ is an $\omega^{q-1}$-constacyclic code over $\mathbb{F}_{q^2}$ of length $n$ with defining set $Z=\bigcup_{j=0}^{\delta}C_{k-(q+1)j}$, where $0\leq \delta\leq (q+1)/4$, then $C^{\bot_H}\subseteq C.$ }\\

\noindent\textbf{Theorem 4.14} \emph{ Let $q$ be an odd prime power with the form $20m+3$ or $20m+7$, where $m$ is a positive integer. Let $n=(q^2+1)/5$, then there exist asymmetric quantum codes with parameters $[[n, n-2(s+t+1), (2s+2)/(2t+2)]]_{q^2}$, where $s, t$ are positive integers and $0\leq t\leq s\leq (q+1)/4.$}\\

\noindent\pf Let $k=(q^2+1)/2$. Suppose $C_2$ is a $q^2$-ary $\omega^{q-1}$-constacyclic code of length $n=(q^2+1)/5$ with defining set $Z_2=\bigcup_{i=0}^{t}C_{k-(q+1)i}$, where $0\leq t\leq (q+1)/4$. Then the dimension of $C_2$ is $n-(2t+1)$. Observe that $Z_2$ consists of $2t+1$ consecutive integers $\{k-(q+1)t,\ldots,k-(q+1),k,k+(q+1),\ldots, k+(q+1)t\}$. From Theorem 2.1, the minimum distance of $C_2$ is at least $2t+2$. From Proposition 2.2, we can see that the minimum distance of $C_2$ is $2t+2$. Hence, $C_2$ is a $q^2$-ary $\omega^{q-1}$-constacyclic code with parameters $[n, n-(2t+1),2t+2]_{q^2}$.

Now, suppose $C_1$ is a $q^2$-ary $\omega^{q-1}$-constacyclic code of length $n=(q^2+1)/5$ with defining set $Z_1=\bigcup_{i=0}^{s}C_{k-(q+1)i}$, where $0\leq t\leq s\leq (q+1)/4$. Similar to the discussion of $C_2$, $C_1$ has parameters $[n, n-(2s+1),2s+2]_{q^2}$. Then from Lemma 2.3 and Theorem 3.2, there exist asymmetric quantum codes with parameters $[[n, n-2(s+t+1), (2s+2)/(2t+2)]]_{q^2}.$ \qed\\

Similar to Theorem 4.14, we have the following result:\\

\noindent\textbf{Theorem 4.15} \emph{ Let $q$ be an odd prime power with the form $20m-3$ or $20m-7$, where $m$ is a positive integer. Let $n=(q^2+1)/5$, then there exist asymmetric quantum codes with parameters $[[n, n-2(s+t+1), (2s+2)/(2t+2)]]_{q^2}$, where $s, t$ are positive integers and $0\leq t\leq s\leq (q-1)/4.$}\\

\noindent\textbf{Remark} From Theorem 4.14, and Theorem 4.15, $d_z+d_x=2s+2t+4$. Then from Theorem 3.3, the constructed asymmetric quantum codes with parameters $[[n, n-2(s+t+1), (2s+2)/(2t+2)]]_{q^2}$ attain asymmetric quantum Singleton bound. Hence, these asymmetric quantum codes are optimal.\\

\noindent\textbf{Example 4.16} {Let $q=23$, then $n=(q^2+1)/5=106$. Suppose the defining set of $\omega^{22}$-constacyclic code $C_1$ is given by $Z_1=C_{265}=\{265\}$. Then $C_1$ is a MDS code with parameters $[106,105,2]_{529}$. We also suppose the defining set of $\omega^{22}$-constacyclic code $C_2$ is given by $Z_2=C_{265}\cup C_{241}\cup C_{217}\cup C_{193}\cup C_{169}\cup C_{145}\cup C_{121}=\{121,145,169,193,217,241,265,289,313,337,361,385,409\}$. Then, $C_2$ is a MDS code with parameters $[106,93,14]_{529}$. From Theorem 4.14, there exists an optimal asymmetric quantum code with parameters $[[106,92,14/2]]_{529}$. By taking the different defining sets of $C_1$ and $C_2$, we can get the optimal asymmetric quantum codes in Table 5.}

\begin{table}[htbp]
{\small
\begin{center}
{\small{\bf TABLE 5}~~Asymmetric quantum codes derived from constacyclic codes of
length 106\\}
\begin{tabular}{c c c c c c c c c c c c c c c c c c c}
\hline
& $[[106,104,2/2]]_{529}$& & &$[[106,100,4/4]]_{529}$ & & &$[[106,94,8/6]]_{529}$& &  &$[[106,86,14/8]]_{529}$&\\
& $[[106,102,4/2]]_{529}$& & &$[[106,98,6/4]]_{529}$ & & &$[[106,92,10/6]]_{529}$ & & &$[[106,88,10/10]]_{529}$&\\
& $[[106,100,6/2]]_{529}$& & &$[[106,96,8/4]]_{529}$ & & &$[[106,90,12/6]]_{529}$& & &$[[106,86,12/10]]_{529}$&\\
& $[[106,98,8/2]]_{529}$& & &$[[106,94,10/4]]_{529}$& & &$[[106,88,14/6]]_{529}$& & &$[[106,84,14/10]]_{529}$&\\
& $[[106,96,10/2]]_{529}$& & &$[[106,92,12/4]]_{529}$& & &$[[106,92,8/8]]_{529}$& & &$[[106,84,12/12]]_{529}$&\\
& $[[106,94,12/2]]_{529}$& & &$[[106,90,14/4]]_{529}$& & &$[[106,90,10/8]]_{529}$& & &$[[106,82,14/12]]_{529}$&\\
& $[[106,92,14/2]]_{529}$& & &$[[106,96,6/6]]_{529}$& & &$[[106,88,12/8]]_{529}$& & &$[[106,80,14/14]]_{529}$&\\
\hline
\end{tabular}
\end{center}}
\end{table}

\noindent\textbf{Example 4.17} {Let $q=17$, then $n=(q^2+1)/5=58$. Suppose the defining set of $\omega^{16}$-constacyclic code $C_1$ is given by $Z_1=C_{145}=\{145\}$. Then $C_1$ is a MDS code with parameters $[58,57,2]_{289}$. We also suppose the defining set of $\omega^{16}$-constacyclic code $C_2$ is given by $Z_2=C_{145}\cup C_{127}\cup C_{109}\cup C_{91}\cup C_{73}=\{73,91,109,127,145,163,181,199,217\}$. Then, $C_2$ is a MDS code with parameters $[58,49,10]_{289}$. From Theorem 4.15, there exists an optimal asymmetric quantum code with parameters $[[58,48,10/2]]_{289}$. By taking the different defining sets of $C_1$ and $C_2$, we can get the optimal asymmetric quantum codes in Table 6.}

\begin{table}[htbp]
{\small
\begin{center}
{\small{\bf TABLE 6}~~Asymmetric quantum codes derived from constacyclic codes of
length 58\\}
\begin{tabular}{c c c c c c c c c c c c c c c c c c c}
\hline
& $[[58,56,2/2]]_{289}$& & &$[[58,52,4/4]]_{289}$ & & &$[[58,46,8/6]]_{289}$&\\
& $[[58,54,4/2]]_{289}$& & &$[[58,50,6/4]]_{289}$ & & &$[[58,44,10/6]]_{289}$&\\
& $[[58,52,6/2]]_{289}$& & &$[[58,48,8/4]]_{289}$ & & &$[[58,44,8/8]]_{289}$&\\
& $[[58,50,8/2]]_{289}$& & &$[[58,46,10/4]]_{289}$& & &$[[58,42,10/8]]_{289}$&\\
& $[[58,48,10/2]]_{289}$& & &$[[58,48,6/6]]_{289}$& & &$[[58,40,10/10]]_{289}$&\\
\hline
\end{tabular}
\end{center}}
\end{table}

\dse{5~~Conclusion}
In this paper we have constructed six new families of asymmetric quantum codes based on constacyclic codes by applying the CSS construction. The new codes achieve the asymmetric quantum Singleton bound  and different from the codes available in the literature. Additionally, the quantum codes constructed in this paper can be utilized in quantum channels with great asymmetry.


\begin{thebibliography}{99}

\bibitem{S95} P. W. Shor, Phys. Rev. A {\bf 52} (1995) 2493.
\bibitem{S96} A. M. Steane, Phys. Rev. Lett. {\bf 77} (1996) 793.
\bibitem{CS96} A. R. Calderbank and P. W. Shor, Phys. Rev. A {\bf 54} (1996) 1098.
\bibitem{CEL99} G. Cohen, S. Encheva and S. Litsyn, IEEE Trans. Inf. Theory {\bf 45} (1999) 2495.
\bibitem{LXW08} Z. Li, L. Xing and X. Wang, Phys. Rev. A {\bf 77} (2008) 012308.
\bibitem{S99} A. M. Steane, IEEE Trans. Inf. Theory {\bf 45} (1999) 1701.
\bibitem{S99a} A. M. Steane, IEEE Trans. Inf. Theory {\bf 45} (1999) 2492.
\bibitem{CLX05} H. Chen, S. Ling and C. Xing, IEEE Trans. Inf. Theory {\bf 51} (2005) 2915.
\bibitem{AK01} A. Ashikhmin and E. Knill, IEEE Trans. Inf. Theory {\bf 47} (2001) 3065.
\bibitem{KKKS06} A. Ketkar, A. Klappenecker, S. Kumar and P. K. Sarvepalli, IEEE Trans. Inf. Theory {\bf 52} (2006) 4892.
\bibitem{AKS07} S. A. Aly, A. Klappenecker and P. K. Sarvepalli, IEEE Trans. Inf. Theory {\bf 53} (2007) 1183.
\bibitem{KZ13} X. S. Kai and S. X. Zhu, IEEE Trans. Inf. Theory {\bf 59} (2013) 1193.
\bibitem{KZ14} X. S. Kai and S. X. Zhu, IEEE Trans. Inf. Theory {\bf 60} (2014) 2080.
\bibitem{G14} G. G. La Guardia, IEEE Trans. Inf. Theory {\bf 60} (2014) 304.
\bibitem{AA08} S. A. Aly and A. Ashikhmin, In: IEEE Information Theory Workshop, pp. 1-5 (2010).
\bibitem{S96} A. M. Steane, Phys. Rev. A {\bf 54} (1996) 4741.
\bibitem{IM07} L. Ioffe and M. Mezard, Phys. Rev. A {\bf 75} (2007) 032345.
\bibitem{SRK08} P. K. Sarvepalli, M. Rotteler and A. Klappenecker, In: Proceedings International Symposium Information Theory, pp. 305-309 (2008).
\bibitem{A08} S. A. Aly, In: Proceedings IEEE International Conference on Computer Engineering and Systems, pp. 157-162 (2008)
\bibitem{SRK09} P. K. Sarvepalli, M. Rotteler and A. Klappenecker,In: Proceedings of the Royal Society A, pp. 1645-1672 (2009).
\bibitem{G11} G. G. La Guardia, Quantum Inf. Comput. {\bf 11} (2011) 0239.
\bibitem{SEDH08} A. M. Stephens, Z. W. E. Evans, S. J. Devitt and L. C. L. Hollenberg, Phys. Rev. A {\bf 77} (2008) 062335.
\bibitem{WFLX10} L. Wang, L. Q. Feng, S. Ling and C. P. Xing, IEEE Trans. Inf. Theory {\bf 56} (2010) 2938.
\bibitem{ELS11} M. F. Ezerman, S. Ling and P. Sol\'{e}, IEEE Trans. Inf. Theory {\bf 57} (2011) 5536.
\bibitem{EL11} M. F. Ezerman and S. Ling, Adv. Math. Commun. {\bf 5} (2011) 41.
\bibitem{CJE11} Y. Chee, S. Jitman and M. F. Ezerman, In: 3rd Int. Castle Meeting on Coding Theroy and Applications, pp. 97-102 (2011).
\bibitem{EJKL13} M. F. Ezerman, S. Jitman, H. M. Kiah and S. Ling, Int. J. Quantum Inf. {\bf 11} (2013) 1350027.
\bibitem{G12} G. G. La Guardia, Quantum Inf. Process {\bf 11} (2012) 591.
\bibitem{Ga12} G. G. La Guardia, Int. J. Quantum Inf. {\bf 10} (2012) 1250005.
\bibitem{G13} G. G. La Guardia, Quantum Inf. Process {\bf 12} (2013) 2771.
\bibitem{G14} G. G. La Guardia, Int. J. Theory Phys. (2014) doi: 10.1007/s10773-014-2031-y.
\bibitem{QZ13} J. F. Qian, L. N. Zhang, Mod. Phys. Lett. B {\bf 27} (2013) 1350010.
\bibitem{CLL14} J. Z. Chen, J. P. Li and J. Lin, Int. J. Theory Phys. {\bf 53} (2014) 72.
\bibitem{ASR01} N. Aydin, I. Siap and D. K. Ray-Chaudhuri, Des. Codes Crypt. {\bf 24} (2001) 313.
\bibitem{MS97} F. J. MacWilliams and N. J. A. Sloane, Amsterdam, The Netherlands: North-Holland, 1997.
\end{thebibliography}
\end{document}